\documentstyle[12pt]{article}
\begin{document}
\title{Dynamics of viscous dissipative gravitational collapse: A full causal approach}
\author{L. Herrera$^1$\thanks{Postal
address: Apartado 80793, Caracas 1080A, Venezuela.} \thanks{e-mail:
laherrera@cantv.net.ve}, A. Di Prisco$^1$\thanks{email: adiprisc@fisica.ciens.ucv.ve}, E. Fuenmayor$^1$\thanks{email: efuenma@fisica.ciens.ucv.ve}, 
\\ and  O. Troconis$^1$\thanks{email: otroconis@fisica.ciens.ucv.ve}\\
{\small $^1$Escuela de F\'{\i}sica, Facultad de Ciencias,}\\
{\small Universidad Central de Venezuela, Caracas, Venezuela.} \\
}
\maketitle 
\newpage
\begin{abstract}
The Misner and Sharp approach to the study of gravitational collapse is extended to the viscous dissipative case in, both, the streaming out and
the diffusion approximations. The dynamical equation is then coupled to  causal transport equations for the heat flux, the shear and the bulk viscosity, in 
the context of Israel--Stewart theory, without excluding the thermodynamics viscous/heat coupling coefficients.  The result is compared with previous works where these later  coefficients were neglected and viscosity variables were not assumed to satisfy causal transport equations. Prospective applications of this result to some astrophysical scenarios are  discussed.

\end{abstract}
\pagebreak
\section{Introduction}
 The gravitational collapse of massive stars represents one of
the few observable phenomena where general relativity is expected to
play a relevant role. This fact is at the origin of  the great attraction that this problem  exerts on the comunity of the relativists,  since  the seminal paper by Oppenheimer
and Snyder \cite{Opp}.

Ever since that work, much was done by
researchers trying to grasp essentials aspects of this phenomenon  (see \cite{May} and references therein).
However this endeavour proved to be difficult and uncertain. Different
kind of obstacles appear, depending on the approach adopted for the
modelling and/or  on the complexity of the physical description of the fluid, assumed to form the selfgravitating object. All these factors in turn are conditioned by   the relevant time scales of different physical phenomena under consideration. 

Thus, during their evolution, self--gravitating objects may pass
through phases of intense dynamical activity, with time scales of the order
of magnitude of (or even smaller than) the hydrostatic time scale, and  for
which the quasi--static
approximation is clearly not reliable, e.g.,the collapse of very massive
stars \cite{8''}, and the quick collapse phase
preceding neutron star formation, see for example \cite{9'} and
references therein. In these cases, in which we are mainly concerned with here, it is mandatory to take into account
terms which describe departure from
equilibrium, i.e. a full dynamic description has to be used. 

We shall assume that the process is dissipative. In fact, it is already an established fact, that gravitational collapse is a
highly dissipative process (see \cite{Hs}, \cite{Hetal}, \cite{Mitra} and references
therein). This dissipation is required to account for the very large
(negative) binding energy of the resulting compact object (of the order
of $-10^{53} erg$).

 Indeed, it appears that the only plausible
mechanism to carry away the bulk of the binding energy of the collapsing
star, leading to a neutron star or black hole is neutrino emission
\cite{1}.

 In the diffusion approximation, it is assumed that the energy flux of
radiation (and that of
thermal conduction) is proportional to the gradient of temperature. This
assumption is in general very sensible, since the mean free path of
particles responsible for the transfer of energy in stellar
interiors is usually very small as compared with the typical
length of the object.
Thus, for a main sequence star as the sun, the mean free path of
photons at the centre, is of the order of $2\, cm$. Also, the
mean free path of trapped neutrinos in compact cores of densities
about $10^{12} \, g.cm.^{-3}$ becomes smaller than the size of the stellar
core \cite{3,4}.

Furthermore, the observational data collected from supernovae 1987A
suggests that the regime of radiation transport prevailing during the
emission process, is closer to the diffusion approximation than to the
streaming out limit \cite{5}.

Dissipative effects in the diffusion approximation are further enhanced by very large values of thermal conductivity, which may be of the order of $\kappa \approx 10^{23} erg \, s^{-1} \, cm^{-1} \, K^{-1}$ for electron conductivity (see \cite{78} and references therein) or even 
$\kappa
\approx 10^{37}erg \, s^{-1} \, cm^{-1} \, K^{-1}$,  for neutrino conductivity in a pre--supernovae event \cite{Ma}.

However in many other circumstances, the mean free path of particles transporting energy may be large enough as to justify the  free streaming approximation. Therefore we
shall include simultaneously both limiting  cases of radiative transport (diffusion and streaming out), allowing for describing a wide range situations. 

Since we are mainly concerned with time scales that might be of the order of (or even smaller than) relaxation times, we have to  appeal to a hyperbolic theory of
dissipation in order to treat the transport equation for dissipative variables. The use of a hyperbolic theory of dissipation is further justified  by the
necessity of overcoming the difficulties inherent to parabolic theories
(see references \cite{Hs}, \cite{18}--\cite{8'} and references therein). Doing so we shall be able to give a description of processes occuring before thermal relaxation.

Some years ago, Misner and Sharp \cite{MisnerSharp} and Misner \cite{Misner} provided a full account of the dynamical equations 
governing the adiabatic \cite{MisnerSharp}, and the
dissipative relativistic collapse in the streaming out approximation \cite{Misner}.

An extension of the Misner dynamical equations as to include dissipation in the form of a radial heat flow (besides pure radiation), was given in \cite{Hs}. In that  work the heat flux was assumed to satisfy a causal transport equation, but  viscosity was absent and thereby  the thermodynamics viscous/heat coupling coefficients were not taken under consideration. Furthermore, for simplification the fluid was assumed shearfree, in despite of the fact  that the relevance of the shear tensor in the evolution of self--gravitaing systems has been brought out by many authors (see  \cite{SFCF} and references therein).

More recently \cite{DHN}, shear viscosity was introduced into the  Misner approach, but  again thermodynamics viscous/heat coupling coefficients were neglected, and furthermore the assumed transport equation for the shear viscosity was the corresponding to the standard Eckart theory of relativistic irreversible thermodynamics \cite{10, 11}.

The motivation to consider viscosity effects in the study of  relativistic gravitational collapse is well founded.  In fact, though they are often excluded in general relativistic
models of stars, they are known to play a very important role in
the structure and evolution of neutron stars. Indeed, depending on
the dominant process, the coefficient of shear viscosity may be as
large as $\eta \approx 10^{20}$ g cm$^{-1}$ s$^{-1}$ (see
\cite{Anderson} for a review on shear viscosity in neutron stars).

On the other hand the coefficient of bulk viscosity may be as
large as $\zeta \approx10^{30}$ g cm$^{-1}$ s$^{-1}$ due to Urca processes in
strange quark matter \cite{sad}. 

Similar and even larger values may be attained for two color superconducting quark matter phases \cite{alford, blaschke} and for hybrid stars \cite{drago}(see also \cite{jones, vandalen, Dong} and references therein for a review
on bulk viscosity in nuclear and quark matter).

The purpose of this work is to present  a dynamical description of the gravitational collapse within the framework of the Misner approach, for the more general  dissipative fluid distribution consistent with spherical symmetry. This includes the presence of both shear and bulk viscosity, with a full causal treatment for all dissipative variables as well as the inclusion of  the thermodynamics viscous/heat coupling coefficients. These coefficients  may be relevant in non--uniform stellar models \cite{8}.

The manuscript is organized as follows:  in the next Section,  besides the field equations, the conventions, and other useful formulae we obtain the resulting dynamical equation.  In Section 3 transport equations  in the context of the
M\"{u}ller--Israel--Stewart theory \cite{Muller67,12,13} are obtained. The coupling of these equations with the dynamical equation  is performed in Section 4. After  doing that we show  how the effective inertial mass density of a  fluid element  reduces by a factor which depends on dissipative variables. This result was already known (see \cite{eim} and references therein), but for the case in which only the heat flux  was assumed to satisfy  a causal transport equation, without the presence of  the thermodynamics viscous/heat coupling coefficients.
 
In Section 5  an expression is derived which relates the Weyl tensor with the density inhomogeneity and thermodynamical variables. This allows us to bring out the role of dissipative variables in a definition of the arrow of time.

 Finally the results are discussed in the  last section.

\section{The basic equations}
 
In this section we shall deploy the relevant equations for describing a viscous dissipative self--gravitating  fluid.  This includes a full description of the matter distribution,
the line element, both, inside and outside
the fluid boundary, and the field equations this line element
must satisfy.

\subsection{Interior spacetime}
We consider a spherically symmetric distribution  of collapsing
 fluid, bounded by a spherical surface $\Sigma$. We
assume the fluid to  undergo dissipation in the
form of heat flow, free streaming radiation and shearing and bulk
viscosity. 

Choosing comoving coordinates inside $\Sigma$, the general
interior metric can be written as
\begin{equation}
ds^2_-=-A^2dt^2+B^2dr^2+(Cr)^2(d\theta^2+\sin^2\theta d\phi^2),
\label{1}
\end{equation}
where $A$, $B$ and $C$ are functions of $t$ and $r$ and are assumed
positive. We number the coordinates $x^0=t$, $x^1=r$, $x^2=\theta$
and $x^3=\phi$.

The assumed matter energy momentum $T_{\alpha\beta}^-$ inside $\Sigma$
has the form
\begin{eqnarray}
T^{-}_{\alpha \beta} = \left( \mu + p + \Pi\right)V_{\alpha}V_{\beta} + (p +
\Pi)g_{\alpha \beta} + q_{\alpha}V_{\beta} + q_{\beta}V_{\alpha} + \epsilon
l_{\alpha}l_{\beta} + \pi_{\alpha \beta} \label{Tmunu}
\end{eqnarray}
where $\mu$ is the energy density, $p$ the  pressure, $\Pi$ the bulk viscosity,
 $q^{\alpha}$ the heat flux, $\pi_{\alpha \beta}$ the shear viscosity tensor, 
$\epsilon$ the radiation density, $V^{\alpha}$ the four velocity of the fluid,
and $l^{\alpha}$ a radial null four vector. These quantities
satisfy
\begin{eqnarray}
V^{\alpha}V_{\alpha}&=&-1, \;\; V^{\alpha}q_{\alpha}=0, \;\;
\;\; l^{\alpha} V_{\alpha}=-1, \;\; l^{\alpha}l_{\alpha}=0,\;\;\nonumber \\
\;\; \pi_{\mu \nu}V^\nu&=&0, \;\;  \pi_{[\mu \nu]}=0, \;\; \pi^\alpha_\alpha=0.
\label{4}
\end{eqnarray}

In the standard irreversible thermodynamics  we have \cite{8, chan1}
\begin{equation}
\pi_{\alpha \beta}=-2\eta \sigma_{\alpha \beta}, \;\; \Pi=-\zeta \Theta
\label{sv}
\end{equation}
where $\eta$ and $\zeta$ denote the coefficient of shear and bulk viscosity, respectively,   $\sigma_{\alpha \beta}$ is the shear tensor and $\Theta$ is the expansion.
However, since we are interested in a full causal picture of dissipative variables we shall not assume (\ref{sv}). Instead, we shall  use the corresponding transport equation derived from the M\"{u}ller--Israel--Stewart theory.

The shear $\sigma_{\alpha\beta}$ is given by
\begin{equation}
\sigma_{\alpha\beta}=V_{(\alpha
;\beta)}+a_{(\alpha}V_{\beta)}-\frac{1}{3}\Theta h_{\alpha\beta}
\label{4a}
\end{equation}
where the acceleration $a_{\alpha}$ and the expansion $\Theta$ are
given by
\begin{equation}
a_{\alpha}=V_{\alpha ;\beta}V^{\beta}, \;\;
\Theta={V^{\alpha}}_{;\alpha} \label{4b}
\end{equation}
and  $h_{\alpha \beta}=g_{\alpha \beta}+V_{\alpha}V_{\beta}$  is the projector 
onto the hypersurface orthogonal to the four velocity .

Since we assumed the metric (\ref{1}) comoving then
\begin{equation}
V^{\alpha}=A^{-1}\delta_0^{\alpha}, \;\;
q^{\alpha}=qB^{-1}\delta^{\alpha}_1, \;\;
l^{\alpha}=A^{-1}\delta^{\alpha}_0+B^{-1}\delta^{\alpha}_1, \;\;
\label{5}
\end{equation}
where $q$ is a function of $t$ and $r$. Also it follows from (\ref{4}) that
\begin{equation}
\pi_{0 \alpha}=0 \;\;,
\pi^1_1=-2\pi^2_2=-2\pi^3_3\;.
\label{5shear}
\end{equation}
In a more compact form we can write
\begin{equation}
\pi_{\alpha \beta}=\Omega(\chi_{\alpha}\chi_{\beta}-\frac{1}{3}h_{\alpha \beta})
\label{comp}
\end{equation}
where $\chi^{\alpha}$ is a unit four vector along the radial direction, satisfying 
\begin{equation}
\chi^{\alpha}\chi_{\alpha}=1, \;\; \chi^{\alpha} V_{\alpha}=0, \;\;\chi^{\alpha}=B^{-1}\delta^{\alpha}_1,
\label{chi}
\end{equation}
 and $\Omega=\frac{3}{2}\pi^1_1$.

With (\ref{5}) we obtain
for (\ref{4a}) its non null components
\begin{equation}
\sigma_{11}=\frac{2}{3}B^2\sigma, \;\;
\sigma_{22}=\frac{\sigma_{33}}{\sin^2\theta}=-\frac{1}{3}(Cr)^2\sigma,
 \label{5a}
\end{equation}
where
\begin{equation}
\sigma=\frac{1}{A}\left(\frac{\dot{B}}{B}-\frac{\dot{C}}{C}\right),\label{5b1}
\end{equation}
and the dot stands for differentiation with respect to $t$, which
gives the scalar quantity
\begin{equation}
\sigma_{\alpha\beta}\sigma^{\alpha\beta}=\frac{2}{3}\sigma^2.
\label{5b}
\end{equation}
For
(\ref{4b}) with (\ref{5}) we have,
\begin{equation}
a_1=\frac{A^{\prime}}{A}, \;\;
\Theta=\frac{1}{A}\left(\frac{\dot{B}}{B}+2\frac{\dot{C}}{C}\right),
\label{5c}
\end{equation}
where the  prime stands for $r$
differentiation.
\subsection{The Einstein equations}
Einstein's field equations for the interior spacetime (\ref{1}) are given by
\begin{equation}
G_{\alpha\beta}^-=8\pi T_{\alpha\beta}^-.
\label{2}
\end{equation}

The non null components of (\ref{2})
with (\ref{1}), (\ref{Tmunu})  (\ref{5}), (\ref{5shear}) and (\ref{comp})
become
\begin{eqnarray}
G^{-}_{00} = 8\pi T^{-}_{00} & =&
8\pi(\mu + \epsilon)A^{2} = \left(2\frac{\dot B}{B} + \frac{\dot C}{C} \right)\frac{\dot
C}{C}\nonumber
\\
&+ & \left(\frac{A}{B} \right)^{2}\left\{-2 \frac{C^{\prime
\prime}}{C} +\left(2 \frac{B^{\prime}}{B} - \frac{C^{\prime}}{C}
\right)\frac{C^{\prime}}{C} + \frac{2}{r}\left(\frac{B^{\prime}}{B}
- 3\frac{C^{\prime}}{C}\right) - \left[ 1 - \left(\frac{B}{C}
\right)^{2}\right]\frac{1}{r^{2}} \right\} \nonumber\\\label{G00}
\end{eqnarray}

\begin{eqnarray}
G^{-}_{11} = 8\pi T^{-}_{11} & =& 8\pi\left[p
+ \Pi + \epsilon + \frac{2}{3} \Omega\right]B^{2} = -\left(\frac{B}{A}
\right)^{2}\left[2\frac{\ddot C}{C} + \left(\frac{\dot C}{C}
\right)^{2} - 2\frac{\dot A}{A}\frac{\dot C}{C}\right] + \nonumber\\
&+& \left(\frac{C^{\prime}}{C} \right)^{2} +
2\frac{A^{\prime}}{A}\frac{C^{\prime}}{C} +
\frac{2}{r}\left(\frac{A^{\prime}}{A} + \frac{C^{\prime}}{C}\right)
+ \left[ 1 - \left(\frac{B}{C} \right)^{2}\right]\frac{1}{r^{2}}
\label{G11}
\end{eqnarray}

\begin{eqnarray}
G^{-}_{22} = 8\pi\ T^{-}_{22}  &=& 8\pi\left[p
+ \Pi - \frac{\Omega}{3} \right](Cr)^{2} = - \left(\frac{Cr}{A}\right)^{2}
\left[\frac{\ddot B}{B} + \frac{\ddot C}{C} - \frac{\dot
A}{A}\left(\frac{\dot B}{B} + \frac{\dot C}{C} \right) + \frac{\dot
B}{B}\frac{\dot C}{C}\right] + \nonumber\\
&+& \left(\frac{Cr}{B}\right)^{2}\left[\frac{A^{\prime \prime}}{A} +
\frac{C^{\prime \prime}}{C} -
\frac{A^{\prime}}{A}\left(\frac{B^{\prime}}{B} -
\frac{C^{\prime}}{C} \right) - \frac{B^{\prime}}{B}
\frac{C^{\prime}}{C} + \frac{1}{r}\left(\frac{A^{\prime}}{A} -
\frac{B^{\prime}}{B} + 2\frac{C^{\prime}}{C}\right)\right]\nonumber\\
\label{G22}
\end{eqnarray}

\begin{eqnarray}
G^{-}_{33} = \sin^2\theta G^{-}_{22}\label{G33}
\end{eqnarray}

\begin{eqnarray}
G^{-}_{01} = 8\pi T^{-}_{01}  = -
8\pi(q + \epsilon )AB =
-2\left(\frac{\dot C^{\prime}}{C} - \frac{\dot
B}{B}\frac{C^{\prime}}{C} - \frac{\dot C}{C}\frac{A^{\prime}}{A}
\right)+ \frac{2}{r}\left(\frac{\dot B}{B} - \frac{\dot C}{C}\right),
\label{G01}
\end{eqnarray}
observe that this last equation may be written as
\begin{equation}
4\pi(q + \epsilon)B=\frac{1}{3}(\Theta-\sigma)^{\prime}-\sigma\frac{(Cr)^{\prime}}{Cr}.
\label{nueva01}
\end{equation}
Next, the mass function $m(t,r)$ introduced by Misner and Sharp
\cite{MisnerSharp}  is defined by

\begin{equation}
m=\frac{(Cr)^3}{2}{R_{23}}^{23}
=\frac{Cr}{2}\left\{\left(\frac{r\dot{C}}{A}\right)^2
-\left[\frac{(Cr)^{\prime}}{B}\right]^2+1\right\}.
 \label{18}
\end{equation}.
\subsection{The exterior spacetime and junction conditions}
Outside $\Sigma$ we assume we have the Vaidya
spacetime (i.e.\ we assume all outgoing radiation is massless),
described by
\begin{equation}
ds^2=-\left(1-\frac{2M(v)}{r}\right)dv^2-2drdv+r^2(d\theta^2
+\sin^2\theta
d\phi^2) \label{1int},
\end{equation}
where $M(v)$  denote the total mass,
and  $v$ is the retarded time.

 The
matching of the full non-adiabatic sphere  (including viscosity) to
the Vaidya spacetime was discussed in
\cite{chan1}. 

From the continuity of the first and second differential forms it follows (see \cite{chan1} for details),
\begin{equation}
m(t,r)\stackrel{\Sigma}{=}M(v), \label{junction1}
\end{equation}
 and 
\begin{eqnarray}
&&2\frac{\dot C^{\prime}}{C}+2\frac{\dot C}{Cr}-2\frac{\dot B}{B}\frac{C^\prime}{C}-2\frac{\dot B}{Br}-2\frac{A^{\prime}}{A}\frac{\dot C}{C} + \nonumber\\&& +\frac{B}{A}\left[2\frac{\ddot C}{C}-2\frac{\dot C}{C}\frac{\dot A}{A}+\left(\frac{A}{Cr}\right)^2+\left(\frac{\dot C}{C}\right)^2-\left(\frac{A}{B}\right)^2 \left(\frac{C^\prime}{C}+\frac{1}{r}\right)\left(\frac{C^\prime}{C}+\frac{1}{r}+2\frac{A^\prime}{A}\right)\right] \nonumber\\&& \stackrel{\Sigma}{=}0,
\label{j2}
\end{eqnarray}
where $\stackrel{\Sigma}{=}$ means that both sides of the equation
are evaluated on $\Sigma$ (observe a misprint in eq.(40) in \cite{chan1} and a slight difference in notation).

Comparing (\ref{j2}) with  (\ref{G11}) and (\ref{G01}) one obtains

\begin{equation}
p+\Pi +\frac{2}{3} \Omega\stackrel{\Sigma}{=}q.
\label{j3}
\end{equation}

Thus   the matching of
(\ref{1})  and (\ref{1int}) on $\Sigma$ implies (\ref{junction1}) and  (\ref{j3}).

In the context of the standard irreversible thermodynamics where (\ref{sv}) is valid, we obtain
\begin{equation}
p+\Pi-\frac{4\eta \sigma}{3}\stackrel{\Sigma}{=}q,
\label{j4}
\end{equation}
which reduces to eq.(41) in \cite{chan1} with the appropriate change in notation. Observe a misprint in eq.(27) in \cite{DHN} (the $\sigma$ appearing there is the one defined in \cite{chan1}, which is $-\frac{1}{3}$ of the one used here and in \cite{DHN}).

\subsection{Dynamical equations}

The non trivial components of the Bianchi identities ,
$(T^{-\alpha\beta})_{;\beta}=0$  yield
\begin{eqnarray}
&&T^{- \mu \nu}_{;\nu}V_{\mu} =
-\frac{1}{A}\left( \dot \mu + \dot \epsilon \right) - \frac{1}{B}\left( q^{\prime} +
\epsilon^{\prime} \right) - 2\left(q
+ \epsilon \right)\frac{(ACr)^{\prime}}{ABCr}
+ \nonumber\\ &&- \frac{2}{A}\frac{\dot C}{C}\left(\mu + p + \Pi +
\epsilon   - \frac{\Omega}{3}
\right) - \frac{1}{A}\frac{\dot B}{B}\left(\mu + p + \Pi + 2\epsilon
 + \frac{2}{3} \Omega \right) \nonumber \\ &&
 = 0 \label{bianchiv}
\end{eqnarray}

and

\begin{eqnarray}
&&T^{- \mu \nu} _{;\nu}\chi_{\mu} =
\frac{1}{A}\left( \dot q + \dot \epsilon \right) +
\frac{2}{A} \frac{(BC)^{.}}{BC}\left( q + \epsilon \right) \nonumber \\ &&+ \frac{1}{B} \left(p^{\prime} + \Pi^{\prime} +
\epsilon^{\prime} + \frac{2}{3} \Omega^{\prime} \right) 
 + \frac{1}{B}
\frac{A^{\prime}}{A}\left(\mu + p + \Pi + 2\epsilon  + \frac{2}{3} \Omega \right) \nonumber\\ &&+
\frac{2}{B}\frac{(Cr)^{\prime}}{Cr}\left(\epsilon +
 \Omega  \right)  = 0. \label{bianchichi}
\end{eqnarray}

To study the dynamical properties of the system, let us  introduce,
following Misner and Sharp \cite{MisnerSharp}, the proper time derivative $D_T$
given by
\begin{equation}
D_T=\frac{1}{A}\frac{\partial}{\partial t}, \label{16}
\end{equation}
and the proper radial derivative $D_R$,
\begin{equation}
D_R=\frac{1}{R^{\prime}}\frac{\partial}{\partial r}, \label{23a}
\end{equation}
where
\begin{equation}
R=Cr \label{23aa}
\end{equation}
defines the proper radius of a spherical surface inside $\Sigma$, as
measured from its area.

Using (\ref{16}) we can define the velocity $U$ of the collapsing
fluid as the variation of the proper radius with respect to proper time, i.e.\
\begin{equation}
U=rD_TC<0 \;\; \mbox{(in the case of collapse)}. \label{19}
\end{equation}
Then (\ref{18}) can be rewritten as
\begin{equation}
E \equiv \frac{(Cr)^{\prime}}{B}=\left[1+U^2-\frac{2m(t,r)}{Cr}\right]^{1/2}.
\label{20}
\end{equation}
With (\ref{23a})-(\ref{23aa}) we can express (\ref{nueva01}) as
\begin{equation}
4\pi(q+\epsilon)=E\left[\frac{1}{3}D_R(\Theta-\sigma)
-\frac{\sigma}{R}\right].\label{21a}
\end{equation}
Observe that in the non--dissipative, shearfree case, the equation above may be written, with the help of (\ref{5b1}), (\ref{5c}) and (\ref{19}) as 
\begin{equation}
D_R(\frac{U}{R})=0
\label{homo}
\end{equation}
implying $U\sim R$, which  describes an homologous collapse \cite{7'}.

Next, using (\ref{G00}-\ref{G01}) and (\ref{16}-\ref{23aa}) we obtain from
(\ref{18})
\begin{eqnarray}
D_Tm=-4\pi R^2\left[\left
(p+\Pi+\epsilon+\frac{2}{3} \Omega \right)U+(q+\epsilon)E\right]
\label{22}
\end{eqnarray}
and
\begin{eqnarray}
D_Rm=4\pi R^2\left[\mu+\epsilon+(q+\epsilon)\frac{U}{E}\right].
\label{27}
\end{eqnarray}
Expression  (\ref{22}) describes the rate of variation of the
total energy inside a surface of radius $R$. On the right hand
side of (\ref{22}), $(p+\Pi+\epsilon+\frac{2}{3} \Omega)U$ (in the case
of collapse $U<0$) increases the energy inside $R$ through the
rate of work being done by the ``effective'' radial pressure
$p+\Pi+ \frac{2}{3} \Omega$  and the radiation pressure $\epsilon$.  In the stationary regime where we can use the standard thermodynamical relation $\pi_{\alpha \beta}=-2\eta \sigma_{\alpha \beta}$, we recover the result obtained in \cite{DHN}.
The second term
$(q+\epsilon)E$ is the matter energy leaving the spherical
surface.

Equation (\ref{27}) shows how the total energy enclosed varies between
neighboring spherical surfaces inside the
fluid distribution.
The first  two terms on the right hand side of (\ref{27}), $\mu+\epsilon$, are due
to the energy density of the fluid element plus the energy density of the
null fluid describing dissipation
in the free streaming approximation. The last term,
$(q+\epsilon)U/E$ is negative (in the case of collapse) and measures the
outflow of heat and radiation. 

The acceleration $D_TU$ of an infalling particle inside $\Sigma$ can
be obtained by using (\ref{G11}), (\ref{18}), (\ref{16})  and (\ref{20}),
producing
\begin{equation}
D_TU=-\frac{m}{R^2}-4\pi R\left(p+\Pi+\epsilon+\frac{2}{3} \Omega\right)
+\frac{EA^{\prime}}{AB}, \label{28}
\end{equation}
and then, substituting $A^{\prime}/A$ from (\ref{28}) into
(\ref{bianchichi}), we obtain 
\begin{eqnarray}
&&\left(\mu + p + \Pi + 2\epsilon + \frac{2}{3} \Omega\right)D_T U=\nonumber \\
&& - \left(\mu + p + \Pi + 2\epsilon + \frac{2}{3} \Omega \right) \left[ \frac{m}{R^2} + 4 \pi R \left(p + \Pi + \epsilon + \frac{2}{3} \Omega \right) \right] \nonumber \\
&& - E^2 \left[  D_R \left(p + \Pi + \epsilon + \frac{2}{3} \Omega \right) + \frac{2}{R} \left( \epsilon +\Omega\right) \right] \nonumber \\
&& - E \left[ D_T q + D_T \epsilon + 4 \left(q + \epsilon \right) \frac{U}{R}+2 \left(q + \epsilon \right) \sigma \right].\label{nd}
\end{eqnarray}

As it can be easily seen, the main difference between (\ref{nd}), and eq.( 40) in \cite{DHN} (regarding the contributions from shear viscosity) stems from the $\pi_{\alpha \beta}$  terms which now are not given by (\ref{sv}), but have to satisfy a transport  equation obtained  within the context of the  causal dissipative theory (see next section). 

Thus, the factor within the round bracket on the left (which equals the factor on the first round bracket on the right, as it should be) represents the effective inertial mass (the passive gravitational mass according to the equivalence principle). 

The first term  on the right hand side of (\ref{nd}) represents the
gravitational force.  In this term, the factor within the  square bracket shows how dissipation 
affects the ``active'' gravitational mass term.  

There are two different contributions in the second square bracket. The
first one is just the gradient of the total ``effective''   pressure
(which includes the radiation
pressure and the influence of shear and bulk viscosity ). The second
contribution comes from the local anisotropy of pressure induced  by  the radiation pressure
and shear viscosity. 

The last square bracket contains different contributions due to dissipative
processes. The third term within this bracket is positive ($U<0$) showing
that the outflow of
$q>0$ and $\epsilon>0$ diminish the total energy inside the collapsing
sphere, thereby reducing the rate of collapse. The last term describes an
effect resulting from the coupling of
the dissipative flux with the shear of the fluid. The effects of
$D_T\epsilon$ have
been discussed  in
\cite{Misner} and we shall not analyze them in detail here.  

Therefore it only remains to analyze the effects of  transport equations when coupled to (\ref{nd});
 we will proceed to carry on that task in the next section.

\section{Transport equations}
As stated in the Introduction, the main purpose of this work consists in providing a full causal description of  viscous dissipative gravitational collapse. This implies that all dissipative variables (noy only $q$)  have to satisfy  the  corresponding transport equation derived from causal thermodynamics. Furthermore the thermodynamics viscous/heat coupling coefficients will not be neglected, as they are expected to be relevant  in non--uniform stellar models \cite{8}.

 Accordingly, we shall use  transport equations derived from the
M\"{u}ller-Israel-Stewart second
order phenomenological theory for dissipative fluids \cite{Muller67, 12, 13}.

This theory was proposed to overcome the pathologies \cite{9} found in the
approaches of Eckart \cite{10} and Landau \cite{11} for
relativistic dissipative processes.  The important point to retain  is that
 this theory provides transport equations for the dissipative variables,  which are  of Cattaneo type \cite{18}, leading
thereby to  hyperbolic equations for  dissipative perturbations.

The starting point  is the general expression for the entropy four--current, which  in the context of the M\"{u}ller-Israel-Stewart theory, reads (see \cite{8} for details)
\begin{equation}
S^\mu=SnV^\mu+\frac{q^\mu}{T}-(\beta_0 \Pi^2+\beta_1 q_\nu q^\nu+\beta_2 \pi_{\nu \kappa}\pi^{\nu \kappa})\frac{V^\mu}{2T}+\frac{\alpha_0 \Pi q^\mu}{T}+\frac{\alpha_1\pi^{\mu \nu}q_\nu}{T}
\label{ent}
\end{equation}
where $n$ is  particle number  density, $\beta_A(\rho,n)$ are  thermodynamic coefficients for  different contributions to the entropy density, and $\alpha_A(\rho, n)$ are thermodynamics viscous/heat coupling coefficients.

Next, from the Gibbs equation and Bianchi identities, it follows that 
\begin{eqnarray}
&&T S^{\alpha}_{;\alpha} = - \Pi \left[ V^\alpha_{; \alpha}-
\alpha_{0}q^{\alpha}_{;\alpha} + \beta_{0} \Pi _{; \alpha} V^\alpha+
\frac{T}{2}\left( \frac{\beta_{0}}{T}V^{\alpha}\right)_{;\alpha}
\Pi\right]  \nonumber\\
&&- q^{\alpha} \left[ h^\mu_{\alpha} (\ln{T })_{,\mu} (1+\alpha_0 \Pi)+ 
V_{\alpha;\mu} V^\mu- \alpha_{0} \Pi_{;\alpha} - \alpha_{1}\pi^{\mu}_{\alpha
; \mu} + \alpha_{1}
\pi^{\mu}_{\alpha}h^\beta_\mu(\ln{T})_{,\beta}\right. \nonumber\\ 
&&\left. + \beta_{1} q_{\alpha;\mu} V^\mu+
\frac{T}{2} \left(
\frac{\beta_{1}}{T}V^{\mu}\right)_{;\mu}q_{\alpha}\right] 
\nonumber\\&& - \pi^{\alpha \mu} \left[ \sigma_{\alpha \mu} -
\alpha_{1}q_{\mu ; \alpha} + \beta_{2} \pi_{\alpha \mu;\nu} V^\nu+
\frac{T}{2} \left(
\frac{\beta_{2}}{T}V^{\nu}\right)_{;\nu}\pi_{\alpha \mu}\right].
\label{diventropia}
\end{eqnarray}\\

Finally, by the standard procedure, the constitutive transport equations follow from the requirement  $S^\alpha;_\alpha \geq 0$
 \begin{eqnarray}
\tau_{0} \Pi_{,\alpha}V^{\alpha} + \Pi =  -\zeta \Theta + \alpha_{0} \zeta
q^{\alpha}_{;\alpha} -  \frac{1}{2}\zeta T\left(
\frac{\tau_{0}}{\zeta T}V^{\alpha}\right)_{;\alpha} \Pi,
\label{ectransppi}
\end{eqnarray}

 \begin{eqnarray}
\tau_{1} h^{\beta}_{\alpha} q_{\beta; \mu}V^{\mu} + q_{\alpha}& =&  - \kappa
\left[ h_{\alpha}^{\beta}T_{, \beta} (1 + \alpha_{0}\Pi) + \alpha_{1}
\pi^{\mu}_{\alpha}h_{\mu}^{\beta}T_{,\beta} + T (a_\alpha - \alpha_{0}
\Pi_{;\alpha} - \alpha_{1}\pi^{\mu}_{\alpha;\mu})\right] \nonumber\\&-&
\frac{1}{2}\kappa T^{2}\left( \frac{\tau_{1}}{\kappa
T^{2}}V^{\beta}\right)_{;\beta} q_{\alpha}\label{ectranspq}
\end{eqnarray}

and 

 \begin{eqnarray}
\tau_{2} h^{\mu}_{\alpha} h^{\nu}_{\beta} \pi_{\mu \nu; \rho}V^{\rho} +
\pi_{\alpha \beta} =  -2\eta \sigma_{\alpha \beta}  + 2\eta
\alpha_{1} q_{<\beta ;\alpha>} - \eta T\left( \frac{\tau_{2}}{2\eta
T}V^{\nu}\right)_{;\nu}\pi_{\alpha \beta}\label{ectransppialphabeta}
\end{eqnarray}

with
\begin{equation}
q_{<\beta;\alpha>}= h^\mu_\beta h^\nu_\alpha \left(\frac{1}{2}(q_{\mu;\nu} + q_{\nu;\mu})- \frac{1}{3} q_{\sigma;\kappa} h^{\sigma \kappa}h_{\mu\nu}\right) \label{marra}
\end{equation}
and where the relaxational times are given by

\begin{equation}
\tau_0=\zeta \beta_0,  \;\; \tau_1=\kappa T \beta_1,  \;\; \tau_2=2\eta \beta_2.
\label{relax}
\end{equation}

Equations (\ref{ectransppi})--(\ref{ectransppialphabeta})  reduce to  equations (2.21), (2.22) and (2.23) in \cite{8} when thermodynamics coupling coefficients vanish.

In our case each of the equations  (\ref{ectransppi})--(\ref{ectransppialphabeta}) has only one independent component, they read
\begin{eqnarray}
\tau_{0} \dot \Pi  &= & - \left(\zeta+  \frac{\tau_{0}}{2} \Pi  \right)A\Theta  + \frac{A}{B} \alpha_{0}
\zeta \left[q^{\prime} +  q \left(
\frac{A^{\prime}}{A}  + \frac{2(rC)^{\prime}}{rC} \right) \right]\nonumber \\
&-&
\Pi \left[\frac{\zeta T}{2} \left(\frac{\tau_{0}}{\zeta T}\right)^{.} + A \right],
\label{resultpi}
\end{eqnarray}

 \begin{eqnarray}
\tau_{1} \dot q &= &-\frac{A}{B} \kappa\{ T^{\prime} (1 + \alpha_{0}\Pi+\frac{2}{3}\alpha_1 \Omega) 
 + T \Big[
\frac{A^{\prime}}{A} - \alpha_{0} \Pi^{\prime} - \frac{2}{3}\alpha_{1} \Big(\Omega^\prime +\Big( \frac{A^{\prime}}{A} + 3\frac{(rC)^{\prime}}{rC} \Big)\Omega \Big)\Big]\}\nonumber \\
&-& q \left[ \frac{\kappa T^{2}}{2}  ( \frac{\tau_{1}}{\kappa
T^{2}})^{.} + \frac{\tau _{1} }{2}  A\Theta + A\right]  \label{resultq}
\end{eqnarray}

and
\begin{eqnarray}
\tau_{2} \dot \Omega =  -2\eta A\sigma + 2\eta
\alpha_{1} \frac{A}{B}\left(q^{\prime} - q \frac{(rC)^{\prime}}{rC}\right)
-\Omega\left[ \eta T  \Big( \frac{\tau_{2}}{2\eta
T} \Big)^{.} + \frac{ \tau_{2}}{2} A \Theta +A\right]. 
\label{ectransppialphabeta11}
\end{eqnarray}\\

We shall now proceed to couple  transport equations in the form above,  to the
dynamical equation (\ref{nd}), in order to bring out
the effects of dissipation  on the dynamics of the collapsing sphere. For
that purpose,  let
us first substitute (\ref{resultq}) into (\ref{nd}),  then we obtain, after some
rearrangements,
\begin{eqnarray}
&&\left(\mu+p+\Pi + 2\epsilon+\frac{2}{3}\Omega\right)(1-\Lambda)D_TU
 =(1-\Lambda)F_{grav}+F_{hyd}\nonumber \\
& +&\frac{\kappa E^2}{\tau_1}\left\{D_RT \left(1+\alpha_0 \Pi +  \frac{2}{3} \alpha_1\Omega\right) -T\left[\alpha_0 D_R \Pi + \frac{2}{3}\alpha_1 \left(D_R \Omega+\frac{3}{R}\Omega\right)\right] \right\}\nonumber\\
&+&E\left[\frac{\kappa T^2q}{2\tau_1}D_T\left(\frac{\tau_1}{\kappa
   T^2}\right)-D_T\epsilon\right]
-E\left[\left(\frac{3q}{2}+2\epsilon\right)\Theta-\frac{q}{\tau_1}-2(q+\epsilon)\frac{U}{R}\right],\nonumber\\
\label{V4N}
\end{eqnarray}
where $F_{grav}$ and $F_{hyd}$ are defined by
\begin{eqnarray}
F_{grav}&=&-\left(\mu+p+\Pi+2\epsilon +\frac{2}{3}\Omega\right)\nonumber\\
&&\times
\left[m+4\pi\left(p+\Pi+\epsilon+\frac{2}{3}\Omega
\right)R^3\right]\frac{1}{R^2},
\label{grav}\\
F_{hyd}&=& -E^2 \left[D_R \left(p+\Pi+\epsilon +\frac{2}{3}\Omega
\right)
+2(\epsilon+\Omega)\frac{1}{R}\right], \label{hyd}
\end{eqnarray}
and $\Lambda$ is given by
\begin{equation}
\Lambda=\frac{\kappa
T}{\tau_1}\left(\mu+p+\Pi+2\epsilon+\frac{2}{3}\Omega\right)^{-1} \left(1-\frac{2}{3}\alpha_1\Omega\right)  .
\label{alpha}
\end{equation}

Next we express  $\Theta$   by means of  (\ref{resultpi}) and feed this back into (\ref{V4N}), obtaining:

\begin{eqnarray}
&&\left(\mu+p+\Pi + 2\epsilon+\frac{2}{3}\Omega\right)(1-\Lambda+\Delta)D_TU
 =(1-\Lambda+\Delta)F_{grav}+F_{hyd}\nonumber \\
& +&\frac{\kappa E^2}{\tau_1}\left\{D_RT \left(1+\alpha_0 \Pi + \frac{2}{3}\alpha_1 \Omega\right) -T\left[\alpha_0 D_R \Pi + \frac{2}{3}\alpha_1 \left(D_R \Omega+\frac{3}{R}\Omega\right)\right] \right\}\nonumber\\
&-&E^2\left(\mu+p+\Pi + 2\epsilon+\frac{2}{3}\Omega\right)\Delta\left(\frac{D_R q}{q}+\frac{2q}{R}\right)
\nonumber\\
&+&E\left[\frac{\kappa T^2q}{2\tau_1}D_T\left(\frac{\tau_1}{\kappa
   T^2}\right)-D_T\epsilon\right]
   +E\left[\frac{q}{\tau_1}+2(q+\epsilon)\frac{U}{R}\right]\nonumber\\
&+&E \frac{\Delta}{\alpha_0\zeta q}\left(\mu+p+\Pi + 2\epsilon+\frac{2}{3}\Omega\right)\left\{\left[1+\frac{\zeta T}{2}D_T\left(\frac{\tau_0}{\zeta T}\right)\right]\Pi + \tau_0 D_T \Pi\right\},\nonumber\\
\label{V4}
\end{eqnarray}
where  $\Delta$ is given by
\begin{equation}
\Delta=\alpha_0 \zeta q \left(\mu+p+\Pi+2\epsilon+\frac{2}{3}\Omega\right)^{-1}\left( \frac{3q+4\epsilon}{2\zeta+\tau_0 \Pi}\right)  .
\label{alpha}
\end{equation}

Thus, once  transport equations have been taken into account, then the
inertial energy density  and the ``passive gravitational mass density'', appear
diminished by the factor $1-\Lambda+\Delta$. This  result generalizes  the  one obtained
 in \cite{DHN}, by means  of  a complete causal treatment of all dissipative variables and the inclusion of  thermodynamics viscous/heat coupling coefficients.

\section{The Weyl tensor}
In this section we shall find some interesting relationships linking
the Weyl tensor with matter variables, from which we shall extract
some conclusions about the arrow of time.

From the Weyl tensor we may construct the Weyl scalar ${\mathcal
C}^2=C^{\alpha\beta\gamma\delta}C_{\alpha\beta\gamma\delta}$ which
can be given in terms of the Kretchman scalar ${\mathcal
R}=R^{\alpha\beta\gamma\delta}R_{\alpha\beta\gamma\delta}$, the
Ricci tensor $R_{\alpha\beta}$ and the curvature scalar R by
\begin{equation}
{\mathcal C}^2={\mathcal
R}-2R^{\alpha\beta}R_{\alpha\beta}+\frac{1}{3}\rm{R}^2.\label{I18}
\end{equation}
With the help of the formulae given in the Appendix of \cite{DHN} and  the field equations, we may write
 (\ref{I18}) as

\begin{equation}
{\mathcal E}=m-
\frac{4\pi}{3}R^3\left(\mu-\Omega \right),\label{I19}
\end{equation}
where ${\mathcal E}$ is given by
\begin{equation}
{\mathcal E}=\frac{{\mathcal C}}{48^{1/2}}R^3. \label{I20}
\end{equation}

From (\ref{I19}) with (\ref{22}) and (\ref{27}) we have
\begin{eqnarray}
D_R{\mathcal E} =4\pi R^2\left[(q+\epsilon)\frac{U}{E}  +\epsilon + \Omega - D_R\left(\mu -\Omega\right)\frac{R}{3}
\right].
\label{II21}
\end{eqnarray}

From (\ref{II21}) we obtain at once for the 
non-dissipative, perfect fluid case
\begin{equation}
D_R{\mathcal E}+\frac{4\pi}{3}R^3D_R\mu=0, \label{arrow}
\end{equation}
implying that $D_R \mu=0$ produces ${\mathcal
C}=0$ (using the regular axis condition), and conversely the
conformally flat condition implies homogeneity in
the energy density.

 Since tidal forces tend to make the gravitating
fluid more inhomogeneous as the evolution proceeds, a relationship like (\ref{arrow})  led Penrose to propose    a gravitational arrow of time in terms of the Weyl tensor \cite{Pe}.

 However the fact that such a relationship is no longer valid in the presence
of local anisotropy of the pressure and/or
dissipative processes, already discussed in \cite{Hetal}, explains its
failure in scenarios where the above-mentioned  factors are present \cite{arrow}.

Here we see how shear viscosity and dissipative fluxes affect the link between
the Weyl tensor and density inhomogeneity. From the above it is evident that  density inhomogeneities may appear in a conformally flat spacetime, if dissipative processes  occur. Examples of this kind have been presented in \cite{anali}.

\section{Conclusions}

We have established the set of equations governing the structure and evolution of self--gravitating spherically symmetric dissipative viscous fluids. 

Dissipative variables  have been assumed to satisfy transport equations derived from  causal thermodynamics, and viscous/heat coupling coefficients have been included. 

As  a result of this aproach we obtain a dynamic equation (\ref{V4}) which shows the influence of dissipative variables and viscous/heat coupling coefficients  on the value of the ``effective'' inertial mass ( ``passive'' gravitational mass). 

In a presupernovae event, dissipative parameters (in particular $\kappa$) may be large enough as to poduce a significative decreasing of the gravitational force term, resulting in a reversal of the collapse. A numerical model exhibiting this kind of ``bouncing'' has been presented in \cite{bounc}.

Nevertheless, the role that this effect  might play in the outcome of gravitational collapse of massive stars will critically depend on specific numerical  values of those quantities.  Such estimations  are, however,  well beyond the scope of this work. 

Here we just want  to display the way those quantities enter into the dynamic equation and stress the fact that they should not be excluded a priori, particularly during the most  rapid phases of the collapse.

From (\ref{II21}) it is apparent that  the production of density inhomogeneities is related to a quantity involving dissipative fluxes and shear viscosity. Thus if
following Penrose we adopt the point of view that self--gravitating systems evolve in the sense of increasing of density inhomogeneity, then any alternative definition for an arrow of time should include those variables.

Finally, it is worth mentioning that we have considered the fluid to be neutral. The reason for this is that, as it can be easily verified, there is not  terms coupling electromagnetic and dissipative variables, in the relevant equations.  Therefore the role of electric charge  in the dynamics of collapse is the same already discussed in \cite{DHN}.

\section{Acknowledgements}
LH wishes to thank FUNDACION EMPRESAS POLAR for financial support and Universitat de les  Illes Balears for financial support and hospitality. ADP  acknowledges hospitality of the
Physics Department of the  Universitat de les  Illes Balears and financial support from the CDCH at Universidad Central de Venezuela. LH and ADP also acknowledge financial support from the CDCH at Universidad Central de Venezuela under grants PG 03-00-6497-2007.

\end{document}